\documentclass[showpacs,aps,prb,amsmath,amssymb,superscriptaddress]{revtex4}
\usepackage{graphicx}
\usepackage{dcolumn}
\usepackage{bm}

\begin{document}
\baselineskip=0.7cm
\title{Ground-state properties of gapped graphene using the random phase approximation}
\author{Alireza Qaiumzadeh}
\affiliation{Institute for Advanced Studies in Basic Sciences
(IASBS), Zanjan, 45195-1159, Iran}
\affiliation{School of physics, Institute for research in fundamental sciences, IPM, 19395-5531 Tehran, Iran}
\author{Reza Asgari}
\affiliation{School of physics, Institute for research in fundamental sciences, IPM, 19395-5531 Tehran, Iran}
\begin{abstract}
We study the effect of band gap on the ground-state properties of Dirac electrons in a doped graphene within the random phase approximation at zero temperature. Band gap dependence of the exchange, correlation and ground-state energies and the compressibility are calculated. In addition, we show that the conductance in the gapped graphene is smaller than gapless one. We also calculate the band gap dependence of charge compressibility and it decreases with increasing the band gap values.
\end{abstract}
%\date{\today}
\maketitle
\section{Introduction}

Graphene is a flat monolayer of carbon atoms tightly packed into a
two-dimensional (2D) honeycomb lattice and it is a basic building block for all
nanostructured carbon. This stable structure has attracted considerable attention because of experimental progress~\cite{novoselov} and because of exotic chiral feature in its electronic properties and
promising applications~\cite{geim}. Very recent experiments on both a suspended graphene and a graphene on substrate have found remarkably high mobility $2\times 10^{5}$ cm$^2 /$Vs for carrier transport~\cite{mobility} at room temperature which is two order of magnitude
higher than the mobility of silicon wafer used in microprocessors.~\cite{dean}

An interesting feature of graphene which makes it very applicable in semiconductor technology is the opening  a gap in the band energy structure of graphene. There are several scenarios to open a gap in the band energy structure of graphene. One is the finite size effect by
using graphene nanoribbons, where gaps give rise to constriction of
the electrons in the ribbon and it depends on the detailed
structure of ribbon edges.~\cite{son} More precisely, the band gaps
with armchair shaped edges originate from quantum confinement
and the value of the gap depends on the width of the ribbon. For
zigzag shaped edges, on the other hand, the band gaps arise from a staggered
sublattice potential due to magnetization at the
edges.~\cite{abanin} Importantly,
edge effects plays a crucial role in transport properties. The gap engineering can be also achieved
through doping the graphene with chemical species~\cite{ohta} due
to the translational symmetry breaking. The electronic
properties of a graphene interacting with {\rm CrO}$_3$ molecules has
been calculated by using {\it ab initio} calculations.~\cite{zanella}
 This type of calculations, predicts opening a gap about $0.12$ eV at the Dirac
point. Another scenario is graphene by placing it on top of an appropriate substrate which breaks the graphene sublattice symmetry and, therefore generates an intrinsic Dirac mass for the charge
carriers.~\cite{substarte} Typical substrate is made of hexagonal $\rm SiC$ with a gap about $0.26$ eV.
A recent band structure calculation for a graphene on top of a hexagonal boron nitride crystal~\cite{giovannetti}
has been shown a band gap about $53$ meV. The gap can also be generated dynamically by
applying a magnetic field.~\cite{gusynin}  Moreover, when both
mono-and bilayer graphene material are covered with water and
ammonia molecules, a gap induce in the spectrum of
energy.~\cite{ribeiro} Interestingly, the mechanism that electrons hopping on a honeycomb lattice with textured
tight-binding hopping amplitudes, the Kekul\`{e} texture, generates a Dirac gap.~\cite{hou} Eventually, It has been suggested that a small gap can be
opened on the Dirac points due to spin-orbit coupling or Rashba effect~\cite{yao}
which makes the system a spin Hall insulator with quantized
spin Hall conductances.~\cite{murakami}

Recently, the local compressibility of
graphene has been measured ~\cite{martin} using a scannable single
electron transistor. The measured
compressibility is claimed to be well described by the kinetic energy
contribution and it is also suggested that exchange and correlation
effects have canceling contributions. From the theoretical point
of view, the compressibility was first calculated by Peres {\it et
al.}~\cite{peres} considering the exchange contribution to a
noninteracting doped or undoped graphene flake. A related quantity
$\partial\mu/\partial n$ (where $\mu$ is the chemical potential
and $n$ is the electron density) is recently considered
by Hwang {\it et al}.\cite{hwang_dmu} within the same approximation and they
stated that correlations and disorder effects would introduce only small corrections.
This statement is only true in quite large density doped values.
To go beyond the exchange contribution, the correlation effects
were taken into account by Barlas
{\it et al.}~\cite{yafis} based on an evaluation of graphene's
exchange and random phase approximation (RPA) correlation
energies. Moreover, Sheehy and Schmalian~\cite{sheehy} by
exploiting the proximity to relativistic electron quantum critical
point, derived explicit expressions for the temperature and
density dependence of the compressibility properties of graphene. Importantly,
the effect of disorder and many-body interactions on the compressibility
has been recently studied by us.~\cite{asgari} We successfully demonstrated the importance of including
correlation effects together with disorder effects in the thermodynamic quantities.
It should be noticed that all these theoretical efforts have been carried out for a gapless graphene.

Our aim in this work is to study the ground-state properties in
the presence of Dirac gap and electron-electron
interactions. For this purpose, we derive the gap dependence of the dynamic polarization function
for a doped graphene to calculate the scattering rate, ground-state energies and the compressibility of the system at
the level of RPA including the opening gap at Dirac point.

We consider different on-site energies for atoms in two sublattices
in graphene which is established experimentally to be important when an appropriate substrate
such a boron nitride or ${\rm SiC}$ is used. The compressibility decreases by increasing the band gap values due to the sublattice symmetry breaking.

The rest of this paper is organized as follows. In Sec.\,II, we
introduce the models for dynamic polarization function and ground-state energy calculations
. We then outline the calculation of d.c conductivity and compressibility.
Section III contains our numerical calculations of ground state
properties. We conclude in Sec.\,IV with a brief summary.

\section{Theoretical Approach}
We consider a Dirac-like electron in a continuum model
interacting via a Coulomb potential $e^2/\epsilon r$ and its
Fourier transform $v_q=2\pi e^2/(\epsilon q)$ where $\epsilon$ is the average background dielectric constant
(for instance, $\epsilon\simeq5.5$ for graphene placed on ${\rm SiC}$ with the other side being exposed to air) having an isotropic band gap at Dirac points.
 If one assumes that
the sublattice symmetry is broken and $\alpha_a$, $\alpha_b$ are on-site energies of atoms $A$ and $B$, respectively, then the contribution of the on-site energies in Hamiltonian of graphene can be written~\cite{gusynin} as

\begin{eqnarray}
\hat{H}_1&=&\sum_{{\bf k},\sigma}[\alpha_aa_\sigma^\dagger({\bf
k})a_\sigma({\bf k})+\alpha_bb_\sigma^\dagger({\bf
k})b_\sigma({\bf k})]\nonumber\\&=&\sum_{{\bf
k},\sigma}\hat{\Psi}^\dagger_{{\bf
k},\sigma}[\alpha_+\tau^0-\alpha_-\tau^3]\otimes\sigma^3\hat{\Psi}_{{\bf
k},\sigma}
\end{eqnarray}
where $\alpha_+=(\alpha_a+\alpha_b)/2$ that corresponds to the same
carrier density on two sublattices,
$\alpha_-=(\alpha_a-\alpha_b)/2$ is carrier imbalance on two
sublattices that leads to break the inversion symmetry and
$\tau^0$ is $2\times2$ unit matrix, $\tau^3$ is a Pauli matrix that acts on $K_+$ and $K_-$
two-degenerate valleys at which $\pi$ and $\pi^*$ bands touch and $\sigma^3$ is Pauli matrix that act on
graphene's pseudospin degrees of freedom.  Consequently, the
noninteracting Hamiltonian for a gapped graphene is given by
$\hat{H}_0=\sum_{{\bf k},\sigma}\Psi^\dagger_{{\bf
k},\sigma}\mathcal{\hat{H}}_0\Psi_{{\bf k},\sigma}$ where
\begin{eqnarray}\label{h0}
\mathcal{\hat{H}}_0=\left(%
\begin{array}{cccc}
  \Delta & \hbar v \hat{k}^* & 0 & 0 \\
  \hbar v \hat{k} & -\Delta & 0 & 0 \\
  0 & 0 & -\Delta & -\hbar v \hat{k}^* \\
  0 & 0 & -\hbar v \hat{k} & \Delta \\
\end{array}%
\right)
\end{eqnarray}
where $\hat{k}=k_x+ik_y$ and $\sigma$ is the spin of charge carrier. Here, $v=3 t a/2$ is the Fermi velocity, $t$ is the tight-binding hopping integral, $a$ is the spacing of the honeycomb lattice. For the hexagonal crystal structure of graphene, $a=1.42${\AA} is the carbon-carbon distance, the tight-bonding hopping energy is $t=2.8$ eV and the bare Fermi velocity is $v=10^6$ $m/s$. In the noninteracting Hamiltonian, $\hat{H}_0$ the reference energy $(\alpha_a+\alpha_b)\tau_0/2$ is subtracted and the energy gap is defined as $2 \Delta=(\alpha_b-\alpha_a)$ where
we expect $\Delta <t$. The corresponding four components pseudospinor of the noninteracting Hamiltonian is $\Psi^\dagger_{{\bf k},\sigma}=\left(%
\begin{array}{cccc}
  \psi^b_{+,\sigma},  \psi^a_{+,\sigma}, \psi^a_{-,\sigma}, \psi^b_{-,\sigma}
\end{array}%
\right)$. It is easy to diagonalize the noninteracting
Hamiltonian based on pseudospinors in the conduction and valance
band of energies with eigenvalues given by $\pm\sqrt{\hbar^2 v^2 k^2+\Delta^2}$. Importantly, the low energy quasiparticle excitations in a gapless graphene are linearly dispersing and it is valid for energy less than 1 eV. Accordingly, the validity of the noninteracting Hamiltonian given by Eq.~(\ref{h0}) to explore graphene properties is to the case which $\sqrt{\hbar^2 v^2 k^2+\Delta^2}< 1 eV$. On the other hand, we shall achieve to a conventional two-dimensional electron gas system by setting $\hbar v k /\Delta \ll 1$.

Finally, the total Hamiltonian including the electron-electron repulsion interaction is given by
\begin{equation}\label{ham}
 \hat{H}= \hat{H}_0+ \frac{1}{2S}\sum_{{\bf
q}\neq 0}v_q ({\hat n}_{\bf q} {\hat n}_{-{\bf q}}-{\hat N}),
\end{equation}
where $S$ is the sample area and ${\hat N}$ is the total number operator. The presence of a neutralizing background of positive charge is
explicit in Eq.~(\ref{ham}). As we mentioned in introduction section, this kind of Hamiltonian can be used in graphene by placing it on top of an appropriate substrate that breaks the graphene sublattice symmetry and generates an intrinsic Dirac gap.

A central quantity in the theoretical formulation of the many-body
effects in Dirac fermions is the noninteracting dynamical polarizability function~\cite{yafis,hwang,others}
$\chi^{(0)}({\bf q},i\Omega,\mu\geq\Delta)$ where $\mu$ is chemical
potential. Here, we would like to emphasize that we have calculated the gap dependence of the noninteracting polarization function for doped graphene however the vacuum polarization function in which $\mu=\Delta$ has been calculated by Kotov {\it et al}~\cite{kotov}. They studied the distribution of polarization charge induced by a Coulomb impurity for undoped graphene. However, we would like to study the ground-state properties for doped graphene sheets. To achieve this goal, we write the dynamical polarizability function in terms of one-body noninteracting
Green's function
\begin{equation}\label{chi0}
\chi^{(0)}({\bf q},\Omega,\mu)=-i\int \frac{d^2{\bf k}}{(2\pi)^2}
\int\frac{d\omega}{2\pi}{\rm Tr}[i\gamma_0 G^{(0)}({\bf k}+{\bf q},
\omega+\Omega,\mu) i\gamma_0 G^{(0)}({\bf k},\omega,\mu)]~,
\end{equation}
where one-body noninteracting Green's function~\cite{chin} by using the noninteracting Hamiltonian
is given by
\begin{equation}
G^{(0)}({\bf k},\omega,\mu)
=i\frac{-\gamma_0\omega+\hbar v{\bf \gamma\cdot k}+i\Delta}{-\omega^2+\hbar^2 v^2k^2+\Delta^2-i\eta}
-\pi\frac{-\gamma_0\omega+\hbar v{\bf \gamma\cdot k }+i\Delta}{\sqrt{\hbar^2 v^2k^2+\Delta^2}}\delta(\hbar\omega-\sqrt{\hbar^2 v^2 k^2+\Delta^2})
\theta(k-{\rm k_F})~,
\end{equation}
in which $\gamma$-matrices are related to Pauli
matrices by $\sigma^3=-i\gamma_0$ and $\sigma^j=(-1)^j \sigma^3 \gamma_j$
for $j=1,2$ and ${\rm k_F}$ is the Fermi momentum related to the density of electron as given by
${\rm k_F}=(4\pi n/g)^{1/2}$. $g=g_v~g_s=4$ is valley and spin degeneracy and $\theta$ is the Heaviside step function. The chemical potential is given by $\mu=\sqrt{\hbar^2 v^2 {\rm k_F}^2+\Delta^2}$ at zero temperature.
After implementing $G^{(0)}({\bf k},\omega,\mu)$ in Eq.~(\ref{chi0})
and calculating the traces and integrals, the result is given by the
follow expression

\begin{eqnarray}\label{eq:final_result}
\chi^{(0)}({\bf q},i \omega,\mu)&=&-\frac{g}{2\pi
v^2}\{\mu- \Delta+\frac{
\varepsilon_q^2}{2}\left[\frac{\Delta}{{\varepsilon_q^2+\hbar^2 \omega^2}}+\frac{1}{2\sqrt{\varepsilon_q^2+\hbar^2\omega^2}}
(1-\frac{4\Delta^2}{\varepsilon_q^2+\hbar^2\omega^2})\tan^{-1}(\frac{\sqrt{\varepsilon_q^2+\hbar^2\omega^2}}{2\Delta})\right]\nonumber\\&-&
\frac{\varepsilon_q^2}{4\sqrt{\hbar^2\omega^2+\varepsilon_q^2}}\Re
e\left [(1-\frac{4\Delta^2}{\varepsilon_q^2+\hbar^2\omega^2})\{\sin^{-1}(\frac{2\mu+i
\hbar\omega}{\varepsilon_q \sqrt{1+\frac{4\Delta^2}{\varepsilon_q^2+\hbar^2\omega^2}}})-\sin^{-1}(\frac{2 \Delta+i
\hbar\omega}{\varepsilon_q \sqrt{1+\frac{4\Delta^2}{\varepsilon_q^2+\hbar^2\omega^2}}})\}\right]\nonumber\\&-&
\frac{\varepsilon_q^2}{4\sqrt{\hbar^2\omega^2+\varepsilon_q^2}}\Re
e\left [(\frac{2\mu+i\hbar\omega}{\varepsilon_q})\sqrt{(1+\frac{4\Delta^2}{\varepsilon_q^2+\hbar^2\omega^2})-(\frac{2\mu+i\hbar\omega}
{\varepsilon_q})^2}\right ]\nonumber\\&+&
\frac{\varepsilon_q^2}{4\sqrt{\hbar\omega^2+\varepsilon_q^2}}\Re
e\left [(\frac{2\Delta+i\hbar\omega}{\varepsilon_q})\sqrt{(1+\frac{4\Delta^2}{\varepsilon_q^2+\hbar^2\omega^2})-(\frac{2\Delta+i\hbar\omega}
{\varepsilon_q})^2}\right ]\}~,
\end{eqnarray}

where $\varepsilon_q=\hbar v q$. By setting $\Delta=0$,
it is easy to determine that Eq.~(\ref{eq:final_result}) reduces to the noninteracting dynamic polarization function of the gapless graphene sheet.~\cite{yafis}  Furthermore, for the half-filed gapped graphene sheet, the noninteracting dynamic polarization function, vacuum polarization, is given by~\cite{kotov}

\begin{equation}
\chi^{(0)}({\bf q},i\omega,\mu=\Delta)=-g\frac{\varepsilon_q^2}{4  v^2\pi}\left[\frac{\Delta}{{\varepsilon_q^2+\hbar^2\omega^2}}
+\frac{1}{2\sqrt{\varepsilon_q^2+\hbar^2\omega^2}}
(1-\frac{4\Delta^2}{\varepsilon_q^2+\hbar^2\omega^2})\tan^{-1}
(\frac{\sqrt{\varepsilon_q^2+\hbar^2\omega^2}}{2\Delta})\right]~.
\end{equation}

Using the above results for the noninteracting polarization function on
the imaginary frequency axis, the density of state at Fermi energy is calculated as
\begin{equation}
D(\varepsilon_F)=D^0(\varepsilon_F)\left [(1+\Delta^2/\varepsilon_{\rm F}^2)^{1/2}\right ]
\theta(\varepsilon_F)~,
\end{equation}
where $D^0(\varepsilon_F)=g \varepsilon_{\rm F}/2\pi
\hbar v^2$ is the density of states of gapless graphene.~\cite{hwang} Note that we define $\varepsilon_{\rm F}=\hbar v {\rm k_F}$.
The linear correction of expanded gapped polarization function for $\Delta$ is zero, however the quadratic correction is easy obtained
\begin{equation}
\chi^{(0)}({q},i\omega,\mu)\simeq\chi^{(0)}({q},i\omega,\mu) \Big{|}_{\Delta=0} -\frac{g}{2\pi v^2}\left[\frac{1}{2}+\frac{\varepsilon_q^2}{\sqrt{\varepsilon_q^2+\omega^2}}\Re e \{ \frac{2\sin^{-1}(\frac{2\mu}{\varepsilon_q})-\pi}{\varepsilon_q^2+\omega^2}- \frac{\sqrt{\varepsilon_q^2-(2+i\omega)^2}}{4\varepsilon_q^2}  \}\right]\Delta^2+O(\Delta^3)~,
\end{equation}

Where the explicit expression of $\chi^{(0)}({q},i\omega,\mu) \Big{|}_{\Delta=0}$ is given by our group.~\cite{yafis} Now, we are in the stage to use the noninteracting polarization function given by Eq.~(\ref{eq:final_result}) to calculate some physical quantities.

\subsection{Transport scattering time in a gapped graphene}

As a first application of the noninteracting polarization function, we would like to calculate the
gapped graphene transport scattering time by randomly distributed impurity centers in
the relaxation time approximation.~\cite{adam} The validity of the Born approximation is discussed by Novikov~\cite{born} and here we use this approximation to calculate qualitatively the graphene transport scattering time. To this purpose,
the transport scattering time is given by
Boltzmann theory,
\begin{eqnarray}
\frac{1}{\tau(\varepsilon_F)}=\frac{2\pi}{\hbar}\sum_{{\bf q}, s,s'}n_i\frac{<|v_i(q)|^2>}{\epsilon( q)^2}
(1-\cos\theta_{{\bf q},{\bf q+{\rm k_F}}})F^{s,s'}({\bf q},{\bf q+{\rm k_F}})
\delta(s \sqrt{\varepsilon_{\rm k_F}^2+\Delta^2}-s'\sqrt{\varepsilon_{{\bf q+{\rm k_F}}}^2+\Delta^2})~,
\end{eqnarray}
where $v_i(q)=\frac{2\pi e^2}{\epsilon q}\exp(-qd)$ is
the Coulomb scattering potential between an electron and an out of
plane impurity, $\epsilon(q)$ is the static RPA dielectric
function appropriate for graphene, $\epsilon(q)=1-v_q\chi^{(0)}(q,0,\mu)$, $n_i$ is the
density of impurities and $d$ is the setback distance from the
graphene sheet and $s,s'$ being $\pm$. Since we consider large charge carrier density and elastic scattering, we can therefore neglect interband scattering process. $F^{\beta}(\bf{q},\bf{q+k}_F)$ is the overlap of states ($\beta=\pm$), which can be easily calculated from the pseudospinors of Hamiltonian, Eq.~(\ref{h0}). The result is as follow
\begin{eqnarray}
F^{\pm}({\bf q},{\bf
q+k})=\frac{1}{2}\left[1\pm\frac{1}{\sqrt{\varepsilon_{k+q}^2+\Delta^2}}\{\sqrt{\varepsilon_k^2+\Delta^2}
+\frac{\varepsilon_q \varepsilon_k cos\phi}{\sqrt{\varepsilon_k^2+\Delta^2}}\}\right]~,
\end{eqnarray}
where $\phi$ is an angle between ${\bf k}$ and ${\bf q}$.
Graphene conductivity can then be calculated by the Boltzmann transport theory with
$\sigma=(e^2/h)2\tau(\varepsilon_{\rm F}) v{\rm k_F}$. The properties of
graphene's Dirac fermions depends on the dimensionless coupling
constant $\alpha_{gr}=g{e^2/\upsilon \epsilon \hbar}$.

\subsection{RPA ground state energy in a gapped graphene}

The ground-state
energies is calculated by using the coupling constant integration
technique, which has the contributions $E_{tot}=E_{kin}+E_{\rm x}
+E_{\rm c}$. The kinetic energy per particle is easy calculated as $ 2\varepsilon_F [(1+\Delta^2/\varepsilon_F^2)^{3/2}-\Delta^3/\varepsilon_F^3]/3$.

The first-order, exchange contribution per particle is given by
\begin{equation}\label{ex}
\varepsilon_{\rm x}=\frac{E_{\rm x}}{N}=\frac{1}{2}\int \frac{d^2
{\bf q}}{(2\pi)^2}~v_q
\left[-\frac{1}{\pi n} \int_0^{+\infty}d
\Omega ~\chi^{(0)}({\bf q},i\omega,\mu)-1\right]\,.
\end{equation}
To evaluate the correlation energy in the RPA, we follow a standard
strategy for uniform continuum models ~\cite{Giuliani_and_Vignale}
\begin{equation}\label{corr}
\varepsilon^{\rm RPA}_{\rm c}=\frac{E_{\rm c}}{N}= \frac{1}{2\pi
n}\int \frac{d^2 {\bf q}}{(2\pi)^2}
\int_0^{+\infty}d\omega\left\{v_q\chi^{(0)}({\bf
q},i\omega,\mu)\right.
\left. +\ln{\left[1-v_q\chi^{(0)}({\bf
q},i\omega,\mu)\right]}\right\}\,.
\end{equation}

Since $\chi^{(0)}({\bf q},i\omega,\mu)$ is linearly
proportional to ${\bf q}$ at large ${\bf q}$ and decrease only like
$\omega^{-1}$ at large $\omega$ in both gapped and gapless graphene, accordingly the exchange and
correlation energy built by Eqs.~(\ref{ex}) and (\ref{corr}) are
divergent.~\cite{yafis, asgari} In order to improve convergence, it is
convenient at this point to add and subtract vacuum polarization, $\chi^{(0)}({\bf
q},i\omega,\mu=\Delta)$, inside the frequency integral and
regularize the exchange and correlation energy. Therefore, these ultraviolet
divergences can be cured calculating
\begin{equation}\label{exchange_regularized}
\delta \varepsilon_{\rm x}=-\frac{1}{2\pi n}\int \frac{d^2 {\bf
q}}{(2\pi)^2}~v_q \int_0^{+\infty}d \omega ~\delta \chi^{(0)}({\bf
q},i\omega,\mu)
\end{equation}
and
\begin{equation}\label{eq:regularization}
\delta \varepsilon^{\rm RPA}_{\rm c}=\frac{1}{2\pi n} \int \frac{d^2
{\bf q}}{(2\pi)^2} \int_0^{+\infty}d\omega\left\{v_q \delta
\chi^{(0)}({\bf q},i\omega,\mu)\right.
\left. +
\ln{\left[\frac{1-v_q\chi^{(0)}({\bf q},i\omega,\mu)}{1-
v_q\chi^{(0)}({\bf q},i\omega,\mu=\Delta)}\right]}\right\}~,
\end{equation}
where $\delta \chi^{(0)}$ is the difference between the doped
($\mu > \Delta$) and undoped ($\mu=\Delta$) polarization functions. With
this regularization, the $q$ integrals have logarithmic ultraviolet
divergences.~\cite{yafis} we can introduce an ultraviolet cutoff
for the wave vector integrals $k_c=\Lambda {\rm k_F}$ which is the order
of the inverse lattice spacing and $\Lambda$ is dimensionless
quantity. Once the
ground-state is obtained the compressibility $\kappa$ can easily
be calculated from $
\kappa^{-1}=n^2\frac{\partial^2 (n \delta\varepsilon_{\rm tot})}
{\partial n^2}~,
$
where the total ground-state energy per particle is given by
$\delta \varepsilon_{\rm tot}=\delta \varepsilon_{\rm kin}+
\delta \varepsilon_{\rm x}+\delta \varepsilon^{\rm RPA}_{\rm c}$.
The compressibility of noninteracting gapless graphene is
$\kappa_0^0=2/(n\varepsilon_F)$ and the compressibility of
noninteracting gapped graphene is given by
$\kappa_0=2/n\varepsilon_F(1+\Delta^2/\varepsilon_F^2)^{1/2}$.

\section{Numerical results}
In this section, we present our calculations for the ground-state
properties of gapped graphene in a continuum model at low energy,
using the model described in the previous section. Our results are considered a n-doped graphene sheet with $\alpha_{\rm gr}=1$ and $2$ where it is a typical value thought to apply to
graphene sheets on the surface of a ${\rm SiC}$ or boron nitride substrates.
We expect that the physics of gapped graphene is different from gapless graphene due to the sublattice symmetry breaking. In the follow, we will investigate these differences, quantitatively.

Fig.~1 shows the noninteracting dynamic polarization function, $\chi^{(0)}({\bf q}, i\omega,\mu)$ of both gapped and gapless graphene in units of the noninteracting density of state at the Fermi surface, $D(\varepsilon_F)$ as functions of $q/{\rm k_F}$ and $\omega/\varepsilon_F$. In both cases,  $\chi^{(0)}({\bf q}, i\omega,\mu)$ linearly diverges with $q$ at small wavelength region and decays as $1/\omega$ at large frequency for finite $\Delta$ values due to interband fluctuations in contrast to the ordinary 2D electron gas.

The function $-\chi^{(0)}({\bf q}, 0,\mu)$ contains a number of noteworthy features is shown in Fig.~2(a). First, as we mentioned before, the $q\rightarrow 0$ limit of the static polarization function is a measure of the number of excited states. Second, the derivative of $\chi^{(0)}({\bf q}, 0,\mu)$ at $q=2 {\rm k_F}$ is singular at finite $\Delta$ values the same as the normal 2D electron gas. Note that $\chi^{(0)}({\bf q}, 0,\mu)$ at $\Delta=0$ is a smooth function. We stress here that the second order correction of the noninteracting polarization function is mostly responsible to this behavior. This singular behavior is responsible for several interesting phenomena such as Friedel oscillations and the associated RKKY interaction.~\cite{Giuliani_and_Vignale} Interestingly, $\chi^{(0)}({\bf q}, 0,\mu)$ reduces to the contentional 2D noninteracting dynamic polarization function at very large $\Delta$ values. In this figure, we have shown $-\chi^{(0)}({\bf q}, 0,\mu)$ at $\Delta=10 \varepsilon_F$ which is exactly the same as the conventional 2D noninteracting dynamic polarization function up to mid $q$ values. The behavior of $\chi^{(0)}({\rm k_F}, i\omega,\mu)$ in unit of the density of state of gapped graphene for various $\Delta$ is displayed in Fig.~2(b).

As an application of the noninteracting polarization function given by Eq.~(\ref{eq:final_result}), we calculate the electric conductivity using the Boltzmann equation. we assume $d=1${\AA} and $\alpha_{\rm gr}=2$. Band gap and density dependence of d.c. conductivity are shown in Fig.~3. Increasing disorder (increasing $n_i$ or
decreasing $d$ for charge-disorder potential ) decrease the $\sigma$ however increasing the gapped value decreases the d.c. conductivity. Our calculations show that $\sigma$ decreases by increasing $\Delta$ as a function of $n/n_i$. Moreover, the density dependence of $\sigma$ is linear at small $d$ and $\triangle$ values and deviates from linearity at large $d$ values.\cite{hwang1} In the inset, we have shown the results for $d=10${\AA} which physically determine that the value of $\sigma$ increases by increasing $d$. Interestingly, a large value of $\sigma$ will be obtained for suspended graphene or by using the ${\rm SiO_2}$ substrate instead of using boron nitride or ${\rm SiC}$ which result in the opening a gap due to symmetry breaking between sublattices.

We also calculated the exchange and correlation energies as a
function of $\Delta$ for various values of the cutoff $\Lambda$. The results are summarized in Fig.~4. we have found that the band gap effects become more appreciable at large
cutoff values. The
exchange energy is positive~\cite{yafis} because our
regularization procedure implicitly selects the chemical potential
of undoped graphene as the half gap energy; doping either occupies
quasiparticle states with energies larger than $\Delta$, or empties
quasiparticles with energies smaller than $-\Delta$. Figure~4(b) shows the
correlation energy $\delta \varepsilon_c$ as a function of
$\Delta$. Note that $\delta
\varepsilon_c$ has the same density dependence as $\delta
\varepsilon_x$ apart from the weak dependence on $\Delta$. In
contrast to the exchange energy, Figure~4(a), the correlation energy is
negative~\cite{yafis}. It is important to note that there is a similar behavior between the kinetic
energy and the exchange-correlation energy as a function of $\Delta$. The kinetic energy is a slowly varying function to $\Delta$ at small gap values and increases by increasing $\Delta$ in middle and large values. Consequently, the total energy increases as a function of $\Delta$. Fig.~5 is shown the total ground-state energy. In the inset, the total energy per particle is shown as a function of $\Delta$
for various values $\alpha_{gr}$ at $\Lambda=50$.

Figure~6 shows the charge compressibility,
$\kappa/\kappa^0_0$ scaled by its noninteracting gapless compressibility as a
function of $\Delta$ for different $\Lambda$ values. The
behavior of $\kappa$ suggests some novel physics
qualitatively different from the physics known in the conventional
2D electron gas.~\cite{yafis, asgari} Kinetic energy and the exchange-correlation energy make negative contributions to the compressibility and therefore reduces the
compressibility by increasing the $\Delta$.

\section{Conclusion}

We have studied the ground-state thermodynamic properties of
a gapped graphene sheet within the random phase approximation (RPA). Note that for a doped graphene the Fermi liquid description is valid. Our aim in this paper is investigating the ground-state properties of a gapped graphene sheet by going from a system with a linear dispersion relation with vanishing the energy gap, $\triangle=0$ to a system with a parabolic dispersion relation where $\triangle\rightarrow\infty$. To achieve this goal, we have calculated the band gap dependence of noninteracting dynamic polarization function for doped graphene sheet. As a consequence, We have presented results for the conductivity suppression over a wide range of energy gap. We have presented results of ground-state energies by incorporating many-body electron-electron interactions via RPA for gapped graphene sheet. The total ground-sate energy increases by increasing the band gap values. This manner occurs based on our model Hamiltonian. We have finally presented results for the charge compressibility suppression over the energy gap. Importantly, the impact of gap energy on the thermodynamic properties would be noticeable for $\triangle \geq 0.2 \varepsilon_{\rm F}$.

Our results demonstrate the importance of including correlation effects together with the gap effects in the thermodynamic quantities of a gapped graphene. It should be possible to extend our work to include disorder effects. Another direction would be to consider the effects of temperature in the thermodynamic quantities.

\begin{acknowledgments}
We would like to thank M. M.
Vazifeh and Kh. Hassani for helpful discussions. A. Q. is supported by IPM grant.
\end{acknowledgments}

\newpage

\begin{figure}[h]
\includegraphics[width=9cm]{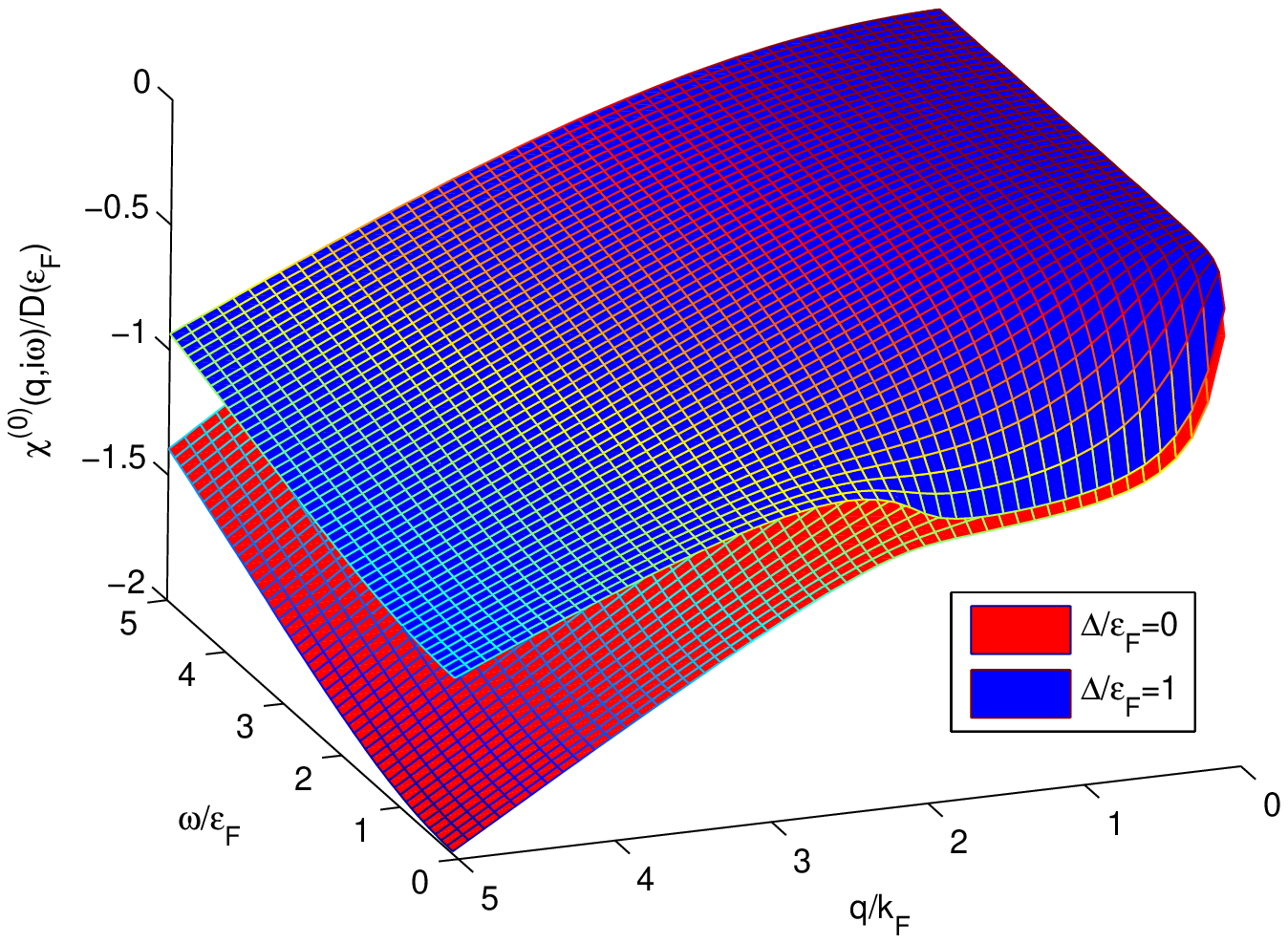}
\caption{(Color online) Noninteracting dynamic polarization function for both gapped and gapless graphene in units of density of state, $D(\varepsilon_{\rm F})$ as functions of $q/{\rm k_F}$ and $\omega/\varepsilon_F$.}
\end{figure}

\begin{figure}[h]
\includegraphics[width=8cm]{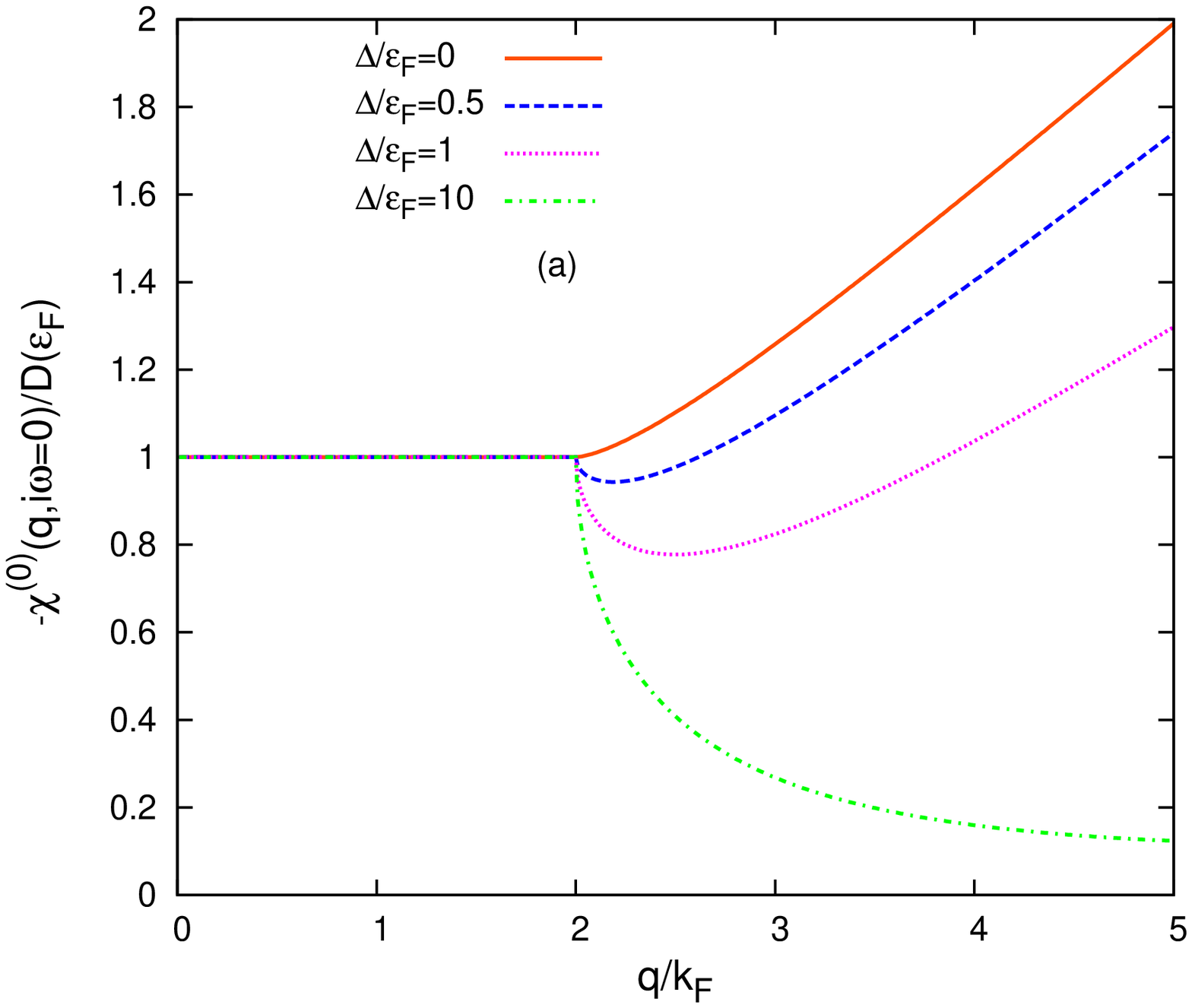}
\includegraphics[width=8cm]{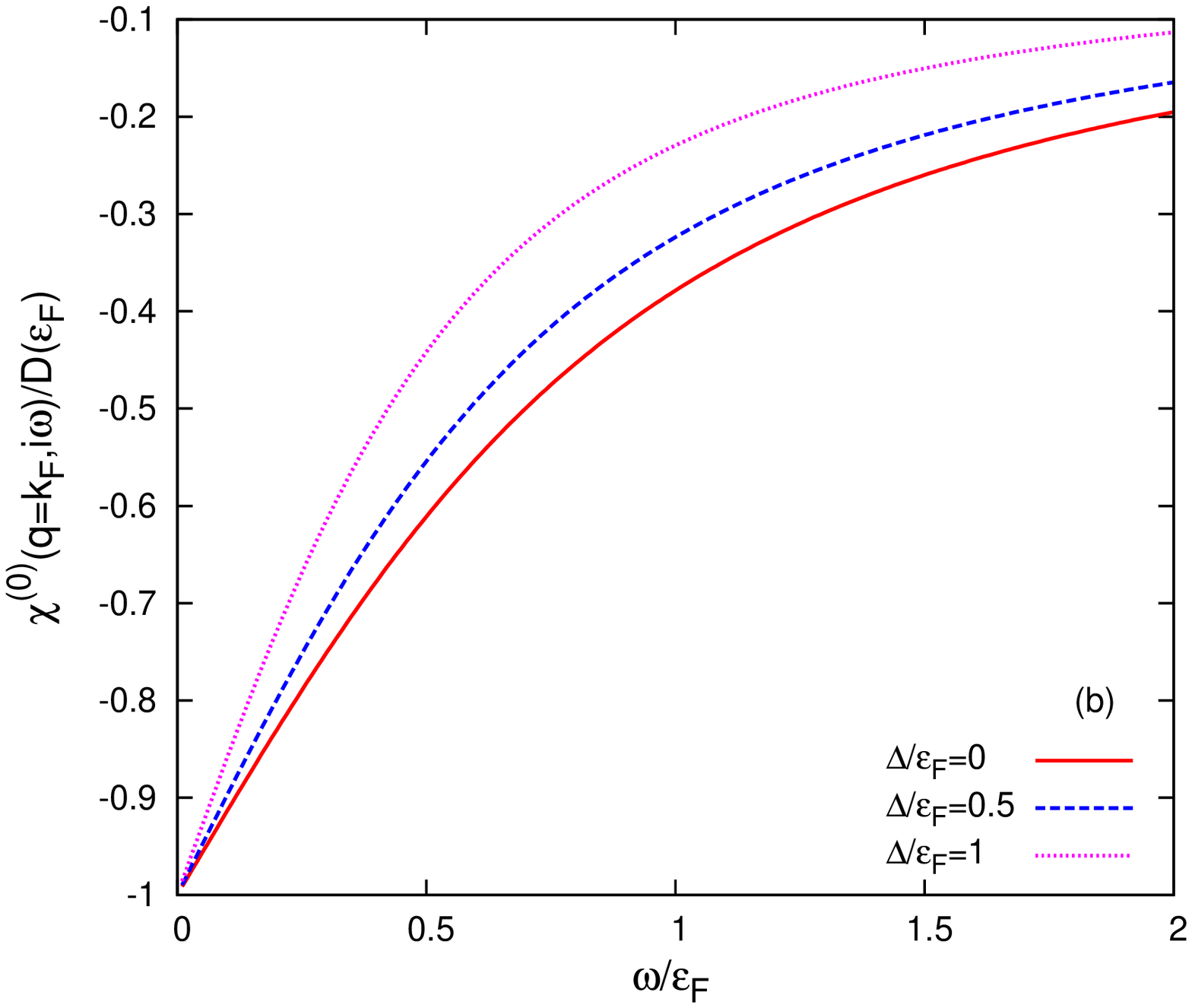}
\caption{(Color online) (a): Static noninteracting polarization function as a function of $q/{\rm k_F}$ for various $\Delta$. (b): $\chi^{(0)}(q=k_{\rm F},i\omega,\mu)$ as a function of $\omega/\varepsilon_F$ for various $\Delta$.}
\label{a}
\end{figure}

\newpage
\begin{figure}[h]
\includegraphics[width=8cm]{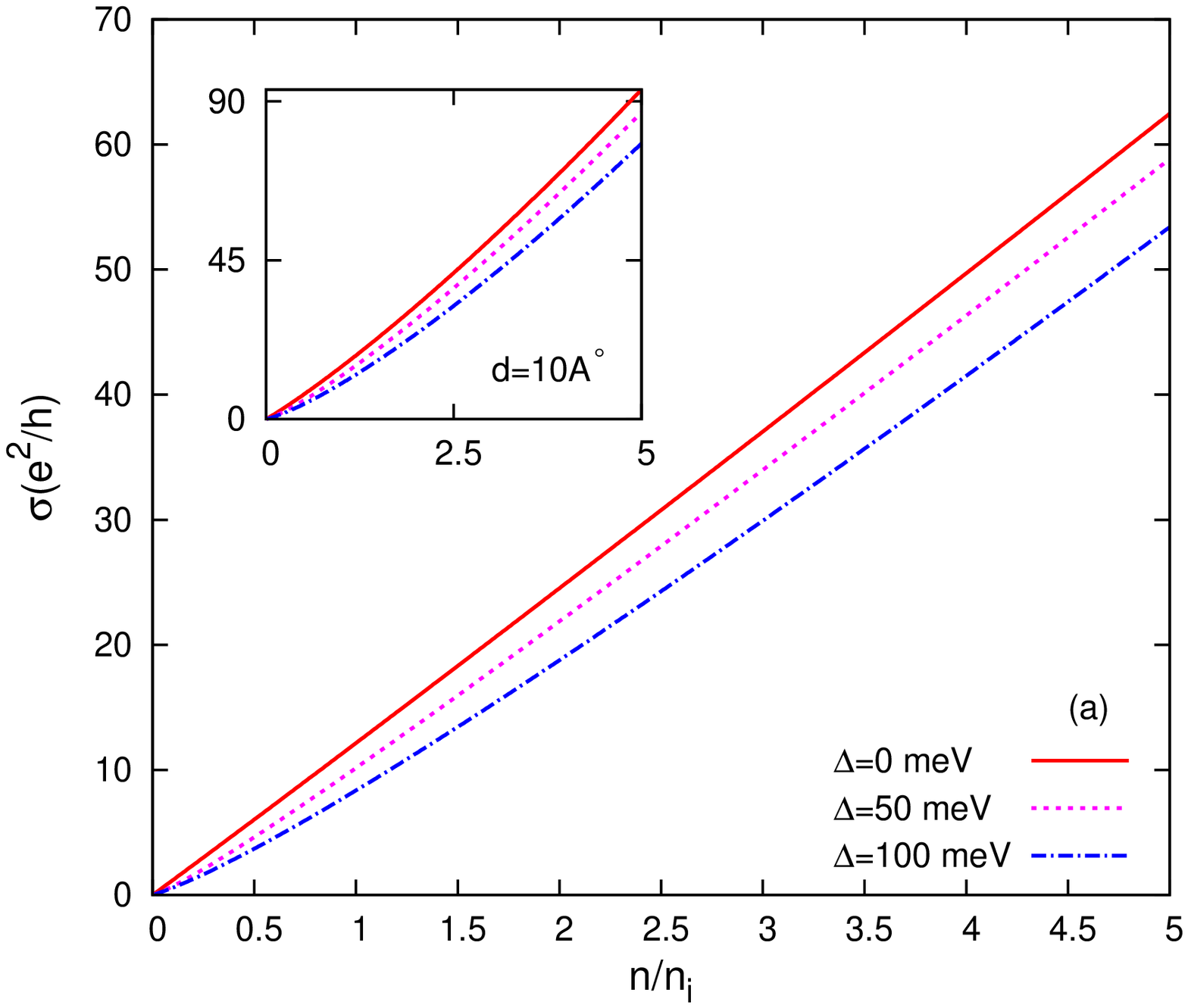}
\includegraphics[width=8cm]{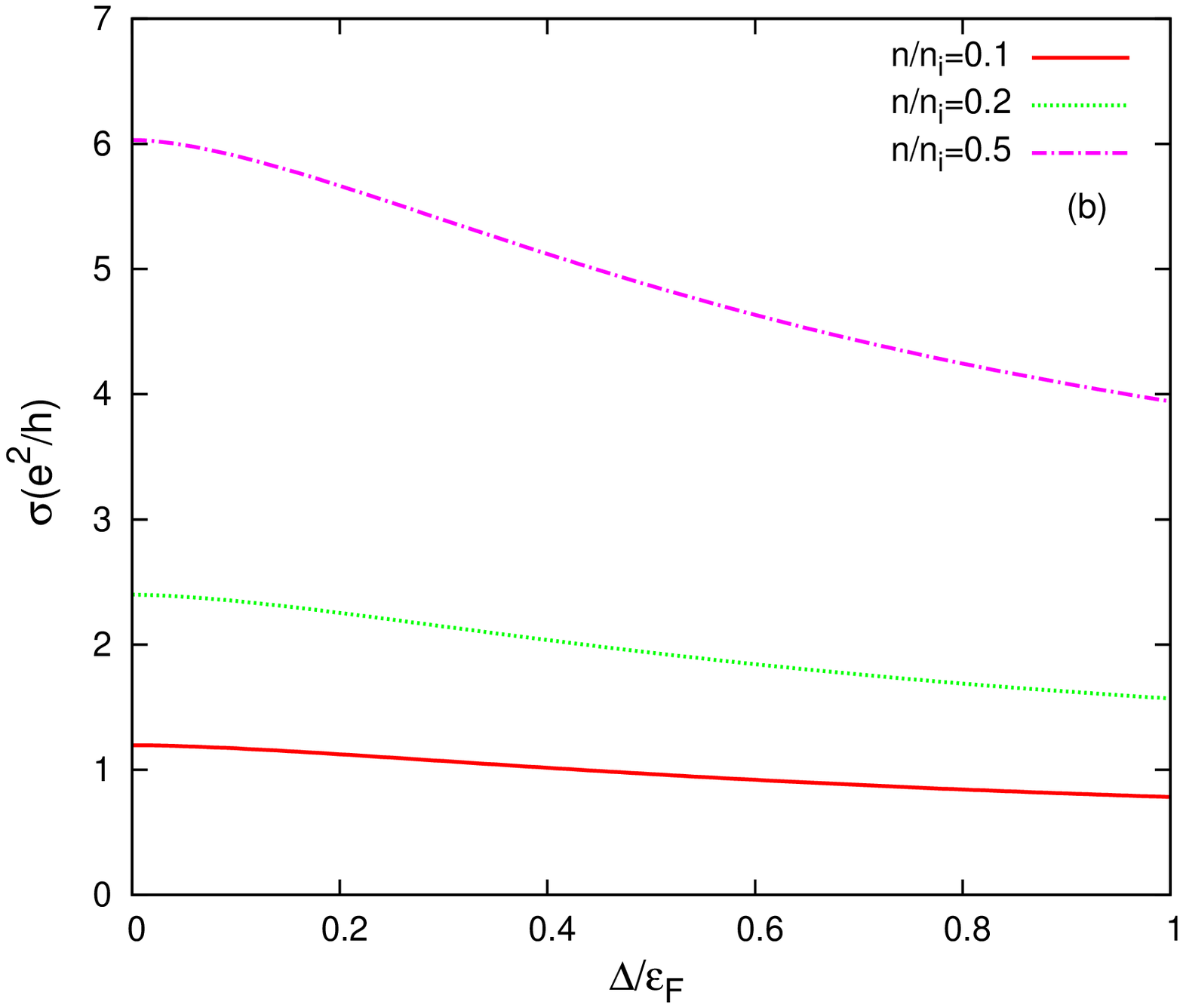}
\caption{(Color online) (a): Conductivity as a function of $n/n_i$ for several
energy gaps at $d=1$ {\AA}, in the inset at $d=10$ {\AA}(b): Conductivity as a function of $\Delta$ for various
electron densities per impurity density. }
\end{figure}

\begin{figure}[h]
\includegraphics[width=8cm]{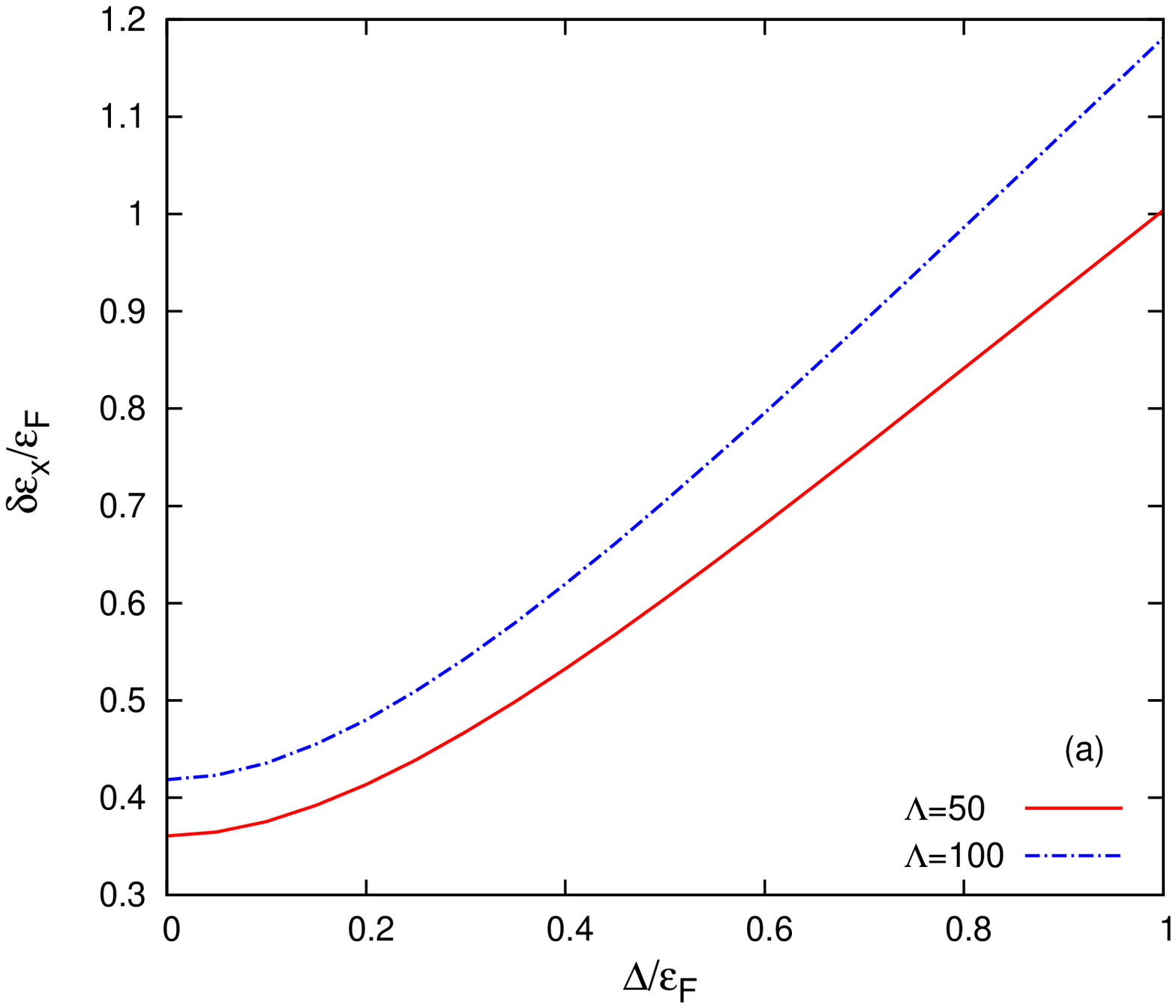}
\includegraphics[width=8cm]{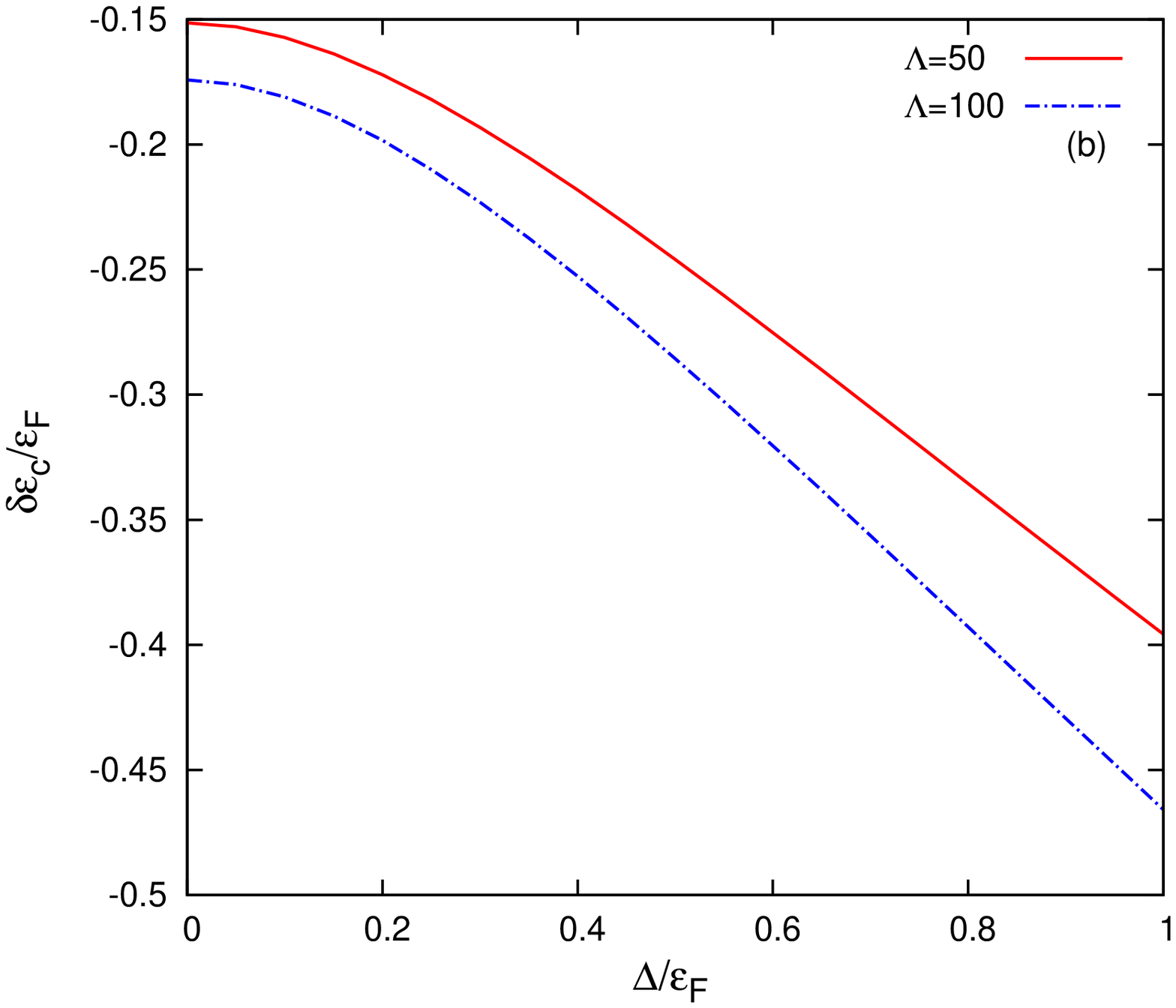}
\caption{(Color online) Exchange (a) and correlation (b) energies as a function of $\Delta$ for various cutoff $\Lambda$ at $\alpha_{gr}=2$.}
\end{figure}

\begin{figure}[h]
\includegraphics[width=9cm]{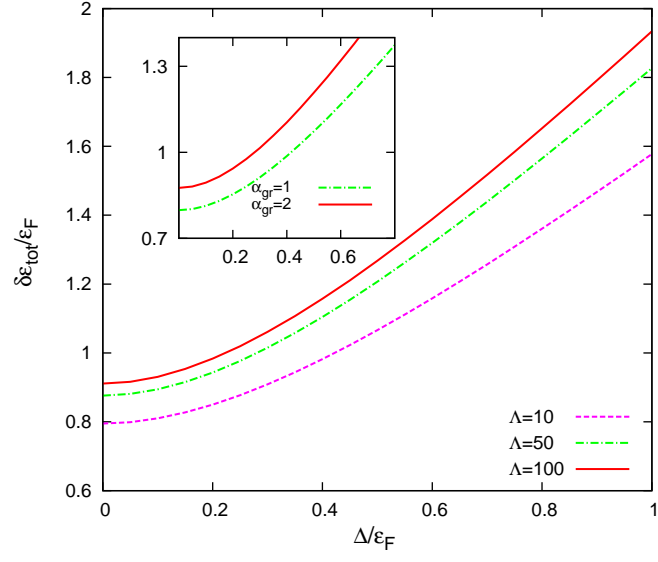}
\caption{(Color online) Total ground-state energy per particle as a function of gap energy, $\Delta$ for various values of the cutoff $\Lambda$ at $\alpha_{gr}=2$. In the inset the total ground-state energy per particle is shown as a function of $\Delta$ for various values of $\alpha_{gr}$ at $\Lambda=50$.}
\end{figure}

\begin{figure}[h]
\includegraphics[width=9cm]{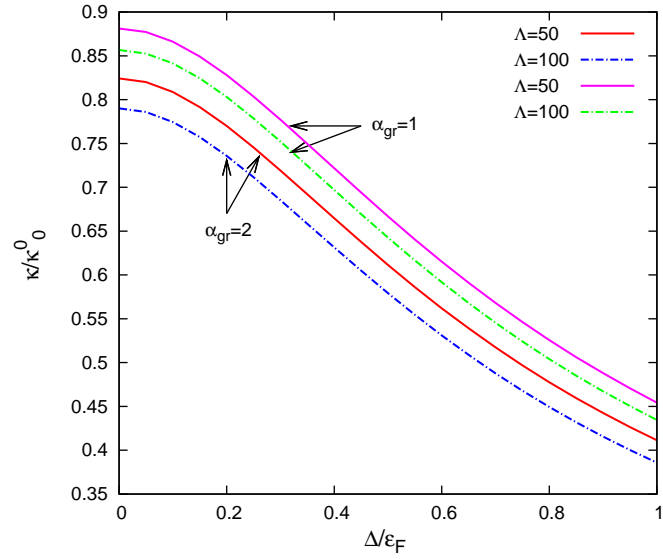}
\caption{(Color online) Compressibility $\kappa/\kappa^0_0$ sacled by that of a noninteracting gapless system as a function of $\Delta$ for various $\alpha_{\rm gr}$ and $\Lambda$. } \label{}
\end{figure}

\end{document}